\documentclass[twocolumn,showpacs,preprintnumbers,
amsmath,amssymb,aps,prd,nofootinbib,superscriptaddress,
eqsecnum]{revtex4}
\usepackage{graphicx,color}
\usepackage[normalem]{ulem}

\renewcommand\sout{\bgroup \color{red} \ULdepth=-.5ex \ULset}
\makeatletter
\renewcommand{\p@subsection}{}
\makeatother

\begin{document}

\title{Implementation of chromomagnetic gluons in Yang-Mills thermodynamics}

\author{Chihiro Sasaki}
\affiliation{%
Frankfurt Institute for Advanced Studies,
D-60438 Frankfurt am Main,
Germany
}

\author{Igor Mishustin}
\affiliation{%
Frankfurt Institute for Advanced Studies,
D-60438 Frankfurt am Main,
Germany
}
\affiliation{%
Kurchatov Institute, Russian Research Center,
Moscow 123182, Russia
}

\author{Krzysztof Redlich}
\affiliation{%
Institute of Theoretical Physics, University of Wroclaw,
PL-50204 Wroc\l aw,
Poland
}
\affiliation{%
Extreme Matter Institute EMMI, GSI,
Planckstr. 1, D-64291 Darmstadt, Germany
}

\date{\today}

\begin{abstract}
Motivated by the recent high-precision lattice data on Yang-Mills
equations of state, we propose an effective theory of SU(3) gluonic matter.
The theory is constructed based on the center and scale symmetries and their
dynamical breaking, so that the interplay between color-electric and
color-magnetic gluons is included coherently. We suggest,  that the magnetic gluon
condensate changes its thermal behavior qualitatively above
the critical temperature, as a consequence of matching to the 
dimensionally-reduced magnetic theories. 
We consider thermodynamics in the mean field
approximation and discuss the consequences for the interaction measure.
\end{abstract}


\maketitle

\section{Introduction}
\label{sec:int}

Non-abelian gauge theories undergo a deconfinement phase transition at finite
temperature $T$. Their bulk asymptotic properties are successfully captured in the
quasi-particle description, which can be consistently calculated in the leading-order
perturbation theory~\cite{blaizot}.
However, a naive perturbative treatment in the weak coupling $g$ is spoiled
since the magnetic screening mass is dynamically generated as a ultra-soft
scale $g^2T$~\cite{Linde,GPY}. The magnetic sector
remains non-perturbative in the high temperature phase, and consequently, the
spatial string tension is non-vanishing for all temperatures~\cite{sst,sst2},
indicating certain confining properties.

This residual interaction brings apparent deviations in equations of state
(EoS) from their Stefan-Boltzmann limit at high temperature. In particular,
the interaction measure $I(T)$ is the best observable to examine dynamical
breaking of scale invariance of the Yang-Mills (YM)
Lagrangian. In lattice simulations of pure SU(3) YM theory the $I(T)/T^2T_c^2$,
with the deconfinement critical temperature $T_c$, is nearly constant in the
range $T_c < T < 5\,T_c$. This observation strongly suggests non-trivial
dynamical effects~\cite{lat,latYM}. Several scenarios have been proposed to explain
this non-perturbative nature, such as, a dimension-2 gluon
condensate that generates
an effective mass term of the gauge boson dynamically~\cite{dim2}, or
a temperature-dependent gluon mass~\cite{CMS,Carter}, as well as
matrix models through introducing an extra $T^2$ term~\cite{fuzzy,matrix}.
Beyond this temperature range the lattice data follow the results
from the Hard Thermal Loop (HTL) resummed perturbation theory.
Thus, a non-perturbative part in the lattice data is extracted by
subtracting the HTL contribution~\cite{latYM}.
The resultant non-perturbative part in
$I(T)/T^2T_c^2$ is {\it monotonically} decreasing, whereas the HTL result
is monotonically increasing with $T$. A plateau that arises in intermediate temperatures
in $I(T)/T^2T_c^2$
can be therefore understood as resulting from the summation of those two contributions.

In this paper,
we formulate an effective theory of SU(3) gluonic matter, which accounts for
two dynamically different contributions,
the chromomagnetic and chromoelectric gluons.

The color-magnetic sector is described by the dilaton, whose condensate
reproduces the trace anomaly of the SU(3) YM theory~\cite{schechter}.
In general, the dilaton
couples also to the Polyakov loop which is the order parameter of
confinement-deconfinement
phase transition and belongs to the color-electric sector.
Thus, the dilaton captures the thermodynamic properties around the critical
point $T_c$, which are related with both, the color-electric and color-magnetic gluons.

Thermal behavior of the magnetic gluon condensate at high temperature is found,
using the three-dimensional YM theories~\cite{AP,Nadkarni,Landsman,Kajantie},
to be $\sim (g^2 T)^4$~\cite{Agasian}. We introduce this contribution to
the effective dilaton potential constructed in four dimensions.
We consider the EoS in this effective theory under the mean field approximation
and discuss the interaction measure and its interpretation.
We also associate our qualitative result with the lattice data.
It turns out that the magnetic gluon condensate can be regarded as
an alternative to the HTL contribution.

\section{Magnetic confinement}
\label{sec:mag}

Color-electric $\langle E\rangle$ and color-magnetic $\langle H\rangle$
gluon condensates behave differently at finite temperature, in particular,
in the deconfined phase~\cite{BElat}. The phase transition is essentially
driven by the electric gluons. The condensate $\langle E\rangle$ drops
toward $T_c$ and approximately vanishes above $T_c$. On the other hand, the
magnetic condensate,
stays nearly constant across the deconfinement phase transition.

Matching the spatial string tension $\sigma_s$, calculated from the
gauge-invariant correlation function of the gauge-field strengths, to
that in the 3-dimensional YM theory, yields the magnetic condensate
as~\cite{Agasian}
\begin{equation}
\langle H \rangle = c_H \left(g^2(T)T\right)^4\,,
\end{equation}
with
\begin{equation}
c_H = \frac{6}{\pi}c_\sigma^2 c_m^2\,.
\end{equation}
The constants $c_\sigma$ and $c_m$ appear in $\sigma_s$ and in the
magnetic gluon mass as
\begin{equation}
\sqrt{\sigma_s(T)} = c_\sigma g^2(T)T\,,
\quad
m_g(T) = c_m g^2(T)T\,.
\end{equation}
For $SU(3)$ YM theory $c_\sigma = 0.566$~\cite{lat} and
$c_m = 0.491$~\cite{3dYM}.

The non-vanishing string tension $\sigma_s$ may support the conjecture,
that in pure YM theory hadronic states, glueballs, can survive in
deconfined phase. The scalar glueballs can be introduced as the dilatons
associated with the scale symmetry. Their condensate
saturates the trace anomaly through the potential~\cite{schechter}
\begin{equation}
V_\chi
= \frac{B}{4}\left(\frac{\chi}{\chi_0}\right)^4
\left[ \ln\left(\frac{\chi}{\chi_0}\right)^4 - 1 \right]\,,
\end{equation}
where $B$ is the bag constant and $\chi_0$ is a dimensionful constant.
The two parameters, $B$ and $\chi_0$, are fixed to reproduce the vacuum energy
density ${\mathcal E} = \frac{1}{4}B = 0.6$ GeV fm$^{-3}$
and the vacuum glueball mass $M_\chi = 1.7$ GeV~\cite{narison,sexton}.
One finds, that $B = (0.368\,\mbox{GeV})^4$ and $\chi_0 = 0.16\,\mbox{GeV}$.

In YM theories, $Z(N_c)$ is
a relevant global symmetry that characterizes the deconfinement
phase transition. The Polyakov loop
$\Phi$
is an order parameter of dynamical breaking of $Z(N_c)$ symmetry~\cite{mclerran}.
The $\Phi$ is introduced as a gauge invariant operator
\begin{eqnarray}
\Phi
&=&
\frac{1}{N_c}\mbox{tr}\hat{L}\,,
\nonumber\\
\hat{L}
&=&
{\mathcal P}\exp\left[i\int_0^{1/T}d\tau A_4(\tau,\vec{x})\right]\,,
\end{eqnarray}
with ${\mathcal P}$ being the Euclidean time ordering and $A_4=iA_0$,
which transforms under $Z(N_c)$   as
\begin{equation}
\Phi \to z\Phi\,,
\quad
z \in Z(N_c)\,.
\end{equation}

The  potential that mixes  the dilaton field and the Polyakov loop should be
manifestly invariant under $Z(N_c)$ and scale transformation.
For $N_c=3$, its most general form is as the following~\cite{Sannino},
\begin{equation}
V_{\rm mix} = \chi^4\left(
G_1\bar{\Phi}\Phi + G_2\left(\bar{\Phi}^3+\Phi^3\right)
{}+ G_3\left(\bar{\Phi}\Phi\right)^2 + \cdots
\right)\,,
\label{mix}
\end{equation}
with unknown coefficients $G_i$.

The Polyakov loop characterizes the chromoelectic
sector of gluons. The dilaton condensate contains the information on both, chromoelectric
and chromomagnetic gluons. 

From the lattice results on those
condensates~\cite{delia}, one concludes,  that the electric component of the dilaton drops
toward the critical point from the side of confined phase.
On the other hand, in deconfined phase, the dilaton represents
chromomagnetic gluo-dynamics.

\section{Effective model}
\label{sec:model}

We formulate the model of gluo-dynamics which accounts for
the interplay between chromoelectric and chromomagnetic gluons as
\begin{equation}
\Omega = \Omega_g + \Omega_\Phi
{}+ V_\chi + V_{\rm mix} + c_0\,.
\label{model}
\end{equation}
The electric gluon part $\Omega_g$ is given
in the presence of a uniform gluon field $A_0$ as~\cite{GPY},
\begin{equation}
\Omega_g
= 2T\int\frac{d^3p}{(2\pi)^3}\mbox{tr}\ln
\left(1 - \hat{L}_A e^{-p/T}\right)\,,
\end{equation}
with the adjoing Polyakol-loop matrix $\hat{L}_A$,
and it can be further expressed in terms of the fundamental Polyakov loop 
$\Phi$ as~\cite{SR}
\begin{equation}
\Omega_g
= 2T \int\frac{d^3p}{(2\pi)^3}\ln
\left( 1 + \sum_{n=1}^8C_n\, e^{-np/T}
\right)\,,
\end{equation}
with
\begin{eqnarray}
C_8
&=&
1\,,
\\
C_1
&=&
C_7
= 1 - 9\bar{\Phi}\Phi\,,
\nonumber\\
C_2
&=&
C_6
= 1 - 27\bar{\Phi}\Phi
{}+ 27\left( \bar{\Phi}^3 + \Phi^3\right)\,,
\nonumber\\
C_3
&=&
C_5
= -2 + 27\bar{\Phi}\Phi
{}- 81\left( \bar{\Phi}\Phi \right)^2\,,
\nonumber\\
C_4
&=&
2\left[
-1 + 9\bar{\Phi}\Phi - 27\left( \bar{\Phi}^3 + \Phi^3\right)
{}+ 81\left( \bar{\Phi}\Phi \right)^2
\right]\,.
\nonumber
\end{eqnarray}

The Haar measure part is introduced as~\cite{fuku}
\begin{equation}
\Omega_\Phi
=
-a_0T\ln\left[ 1 - 6\bar{\Phi}\Phi + 4\left( \Phi^3 + \bar{\Phi}^3\right)
{}- 3\left(\bar{\Phi}\Phi\right)^2\right]\,.
\end{equation}

To formulate an effective mixing between the Polyakov loop and  dilaton,
we take only the first term of Eq.~(\ref{mix}).
Thus,
\begin{equation}
V_{\rm mix}
= G \left(\frac{\chi}{\chi_0}\right)^4 \bar{\Phi}\Phi\,.
\end{equation}
In general, the coupling $G$ can be temperature dependent,
but we consider $G$ as a constant and fix its value to reproduce
the expectation value $\langle\Phi\rangle =0.4$ at $T_c$.
Requiring that a first-order phase transition appears at $T_c=270$ MeV
as found in the lattice results \cite{lat},
one finds that
$a_0 = (0.184\,\mbox{GeV})^3$, $c_0 = (0.244\,\mbox{GeV})^4$
and $G = (0.206\,\mbox{GeV})^4$.

Under the mean field approximation, the temperature dependence of $\langle\Phi\rangle$  and
$\chi$ are obtained from the stationary  conditions for the effective potential
(\ref{model}), $\partial\Omega/\partial\Phi=\partial\Omega/\partial\bar{\Phi}
=\partial\Omega/\partial\chi = 0$,
resulting in coupled gap equations~\footnote{
  We note,  that  $\langle\bar{\Phi}\rangle= \langle{\Phi}\rangle$.
}.
The gap equation for $\langle{\Phi}\rangle$  is solved numerically, whereas  that
for $\chi$ can be solved analytically as
\begin{equation}
\langle\chi\rangle
= \chi_0 \exp\left[-G\langle\bar{\Phi}\Phi\rangle/B\right]\,.
\label{chiT}
\end{equation}

\begin{figure}
\begin{center}
\includegraphics[width=8cm]{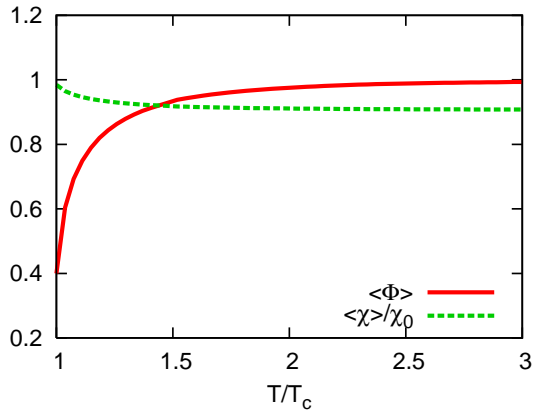}
\caption{
Thermal expectation values of the Polyakov loop (solid) and
the dilaton field (dashed).
}
\label{fig:cond}
\end{center}
\end{figure}

Fig.~\ref{fig:cond} shows the expectation values of $\Phi$ and $\chi$ as the
solutions of the gap equations.
The thermal effect of $\langle\chi\rangle$ is induced via the mixing to the Polyakov loop,
which exhibits a rather weak dependence on temperature above $T/T_c \sim 1.5$.

At higher temperature, due to the dimensional reduction, the theory
in four dimensions should match the three-dimensional YM theory.
We postulate the following matching condition,
\begin{equation}
\frac{\langle\chi\rangle}{\chi_0}
= \left(\frac{\langle H \rangle}{H_0}\right)^{1/4}\,,
\label{magnetic}
\end{equation}
which at a certain temperature $T_{\rm match}$,  should be met with Eq.~(\ref{chiT}).

A constant $H_0$ in Eq. (\ref{magnetic})  is chosen such, that
the model reproduces the 30 \% reduction of the non-perturbative contribution
to the interaction measure $I(T)/T^2T_c^2$ at the matching temperature to the HTL result,
as observed in the lattice calculation~\cite{latYM}.
This implies that $H_0 = (0.8\,\mbox{GeV})^4$~\footnote{
 In fact, for a typical temperature under consideration, this is a natural scale
 compatible to $\sim g^2T$ with the running
 coupling (\ref{running}).
}.

\begin{figure}
\begin{center}
\includegraphics[width=8cm]{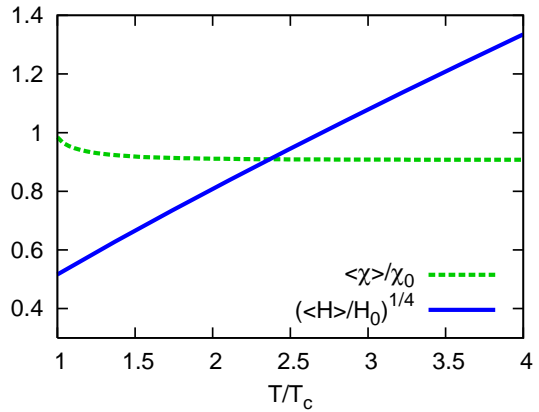}
\caption{
Thermal expectation value of the dilaton (dashed) and
the magnetic condensate (solid).
}
\label{fig:H}
\end{center}
\end{figure}

The matching temperature $T_{\rm match}$ can be extracted from a comparison
of Eqs.~(\ref{chiT}) and (\ref{magnetic}). Applying the two-loop running
coupling,
\begin{eqnarray}
\label{running}
g^{-2}(T)
&=&
2b_0\ln\frac{T}{\Lambda_\sigma}
{}+ \frac{b_1}{b_0}\ln\left(2\ln\frac{T}{\Lambda_\sigma}\right)\,,
\nonumber\\
b_0
&=&
\frac{11}{16\pi^2}\,,
\quad
b_1 = \frac{51}{128\pi^2}\,,
\end{eqnarray}
with $\Lambda_\sigma = 0.104\,T_c$~\cite{lat}, one finds, that
$T_{\rm match} \sim 2.4\,T_c$ (see Fig.~\ref{fig:H}).

In the present model, the  changeover  in  the temperature dependence of the
magnetic condensate, seen in Fig.~\ref{fig:H},  appears by construction.
In YM thermodynamics, however, such behavior emerges dynamically.
A qualitative change of the thermal gluon behavior can be indeed
seen in the spatial string tension at  $T \sim 2\,T_c$~\cite{Agasian}.
Clearly, this property of the magnetic condensate should also affect other thermodynamic quantities, such as the interaction measure
which is considered in the next section.

\section{Trace anomaly and the magnetic scaling}
\label{sec:eos}

We focus on the interaction measure, $I = {\mathcal E} - 3P$,
of the SU(3) YM theory defined in terms of
the pressure $P$ and energy density ${\mathcal E}$.
If the entire interaction vanishes, the interaction measure should
vanish as well,  since the system becomes conformal. However, due to the magnetic
confinement of the pure YM theories, one expects, that even at very high temperature,
$T \gg T_c$, the interaction measure is non-vanishing.
Generally, the $I(T)$ can be parameterized as
\begin{equation}
\frac{I}{T^4}
=
\frac{\mathcal A}{T^2} + \frac{\mathcal B}{T^3}
{}+ \frac{\mathcal C}{T^4} + {\mathcal D}\,.
\label{paramet}
\end{equation}
The model~(\ref{model}) yields the coefficients as
\begin{eqnarray}
{\mathcal A}
&=&
0\,,
\nonumber\\
{\mathcal B}
&=&
\frac{3\Omega_\Phi}{T}\,,
\nonumber\\
{\mathcal C}
&=&
4\left( V_\chi + V_{\rm mix} \right)\,,
\nonumber\\
{\mathcal D}
&=&
\frac{1}{T^4}\left(
3\Omega_g -2\int\frac{d^3p}{(2\pi)^3}
\frac{p\sum_{n=1}^8 n\,C_n\,e^{-np/T}}{1 + \sum_{n=1}^8 C_n e^{-np/T}}
\right)\,.
\nonumber\\
\label{int}
\end{eqnarray}
Obviously, the last term ${\mathcal D}$ should survive
at high $T$, so that the interaction measure still exists. However, this 
contribution approaches zero,  already at moderate temperature,  
since $\langle\Phi\rangle\to 1, $
as seen
in Fig.~\ref{fig:cond}. Indeed one finds, that
\begin{eqnarray}
{\mathcal D}
&\stackrel{\langle\Phi\rangle \to 1}{\to}&
\frac{3\cdot16}{T^3}\int\frac{d^3p}{(2\pi)^3}\ln\left(1-e^{-p/T}\right)
\nonumber\\
&&
{}+ \frac{16}{T^4}\int\frac{d^3p}{(2\pi)^3}\frac{pe^{-p/T}}{1-e^{-p/T}}
\nonumber\\
&=&
8\pi^2\left( \frac{-1}{15} + \frac{1}{15}\right)\,
= 0\,.
\label{d:model}
\end{eqnarray}

Requirement of the non-vanishing ${\mathcal D}$ can be discriminated in $I/T^2$
since it appears as a coefficient of the quadratic,  $T^2$ term, whereas other
contributions with ${\mathcal B}$ and ${\mathcal C}$ monotonically decrease.
From Eq.~(\ref{d:model}), one finds,  that any residual contribution $\sim T^2$ does
not show up in $I/T^2$. Recall, that the coefficient ${\mathcal D}$ in Eq.~(\ref{int})
is entirely chromoelectric,  since it does not contain $\chi$. Therefore,
in order to introduce magnetic confinement effectively which yields
residual interaction at high temperature, such that $I \neq 0$, one transmutes
$\chi$
into $\langle H\rangle \sim (g^2(T)T)^4$ via
Eq.~(\ref{magnetic}), and applies it to Eq.~(\ref{model}).
This generates a $T^4$ contribution,  which appears
from the $V_\chi + V_{\rm mix}$ part, and results in the equations of state
deviating from the Stefan-Boltzmann values at high temperature.
One also finds an additional contribution
to the interaction measure from $\langle H\rangle$,  as
\begin{equation}
\delta I
= -B\frac{\langle H\rangle}{H_0}
{}+ \left(2b_0 + \frac{b_1}{b_0}
\frac{1}{\ln\left(T/\Lambda_\sigma\right)}\right)
\frac{\langle H\rangle}{g^4(T)H_0}\,.
\label{delta}
\end{equation}
The first term is of order ${\mathcal O}(g^8)$, whereas the second
is ${\mathcal O}(g^4)$ which is thus the leading contribution to $I$.

\begin{figure}
\begin{center}
\includegraphics[width=8cm]{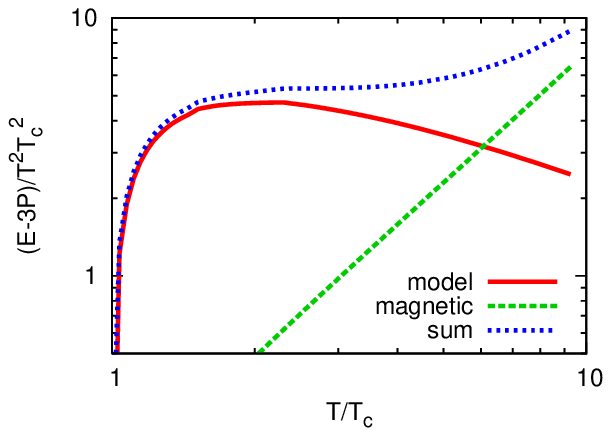}
\caption{
The interaction measure normalized by $T^2 T_c^2$.
The dashed curve labeled with ``magnetic'' corresponds to
$\delta I$ given in Eq.~(\ref{delta}).
}
\label{fig:intT2}
\end{center}
\end{figure}

The interaction measure normalized by $T^2T_c^2$ is shown in
Figs.~\ref{fig:intT2}.
The $I/T^2T_c^2$ is monotonically decreasing even at high temperature when
no matching to the 3-dim YM is made.
The magnetic contribution generates a $T^2$ dependence,  as seen in the figure.
The sum of those two contributions forms  a plateau-like behavior 
in    $I/T^2T_c^2$
at moderate temperature, $T/T_c \sim 2$-$4$.
This property appears due to the residual chromomagnetic interaction 
encoded in the dilaton, $\chi^4 \sim H$.
The resulting behavior of $I/T^2T_c^2$ with temperature,
seen in Fig.~\ref{fig:intT2}, qualitatively agrees
with the latest high-precision lattice data~\cite{latYM}.
We note that a smooth switching from the dilaton to the magnetic condensate
must happen dynamically, so that thermodynamic quantities, such as the
specific heat, do not experience any irregular behavior above $T_c$.

\begin{figure}
\begin{center}
\includegraphics[width=8cm]{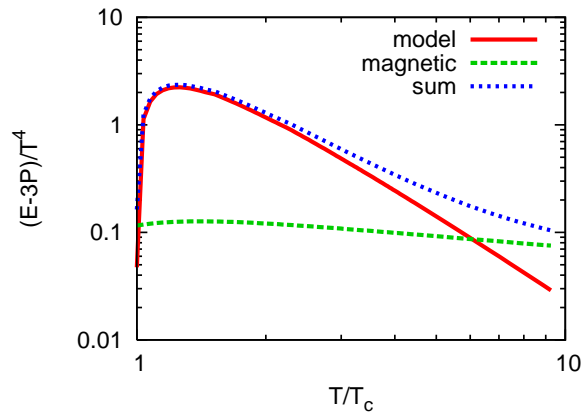}
\caption{
The interaction measure normalized by $T^4$.
The dashed curve labeled with ``magnetic'' corresponds to
$\delta I$ given in Eq.~(\ref{delta}).
}
\label{fig:intT4}
\end{center}
\end{figure}

Fig.~\ref{fig:intT4} shows the interaction measure normalized by $T^4$.
With such normalization, the impact of the magnetic contribution is not
well distinguishable. Thus, in order to identify different dynamical
effects contributing to the interaction measure, it is indeed more 
appropriate to normalize
$I(T)$ by $T^2T_c^2$, as was suggested in~\cite{matrix}.

The lattice data have  confirmed also,  that there is a non-vanishing $T^2$
contribution to the interaction measure $I(T)$~\cite{latYM}.
This would correspond to a non-vanishing ${\mathcal A}$ in Eq.~(\ref{paramet}).
Such a term can appear from a dynamically generated gluon mass.
Imposing the scale symmetry, the expected mass
term is of the following form~\cite{Carter},
\begin{equation}
{\mathcal L}_m
= \frac{1}{2}G_A^2\left(\frac{\chi}{\chi_0}\right)^2 A_\mu A^\mu\,,
\end{equation}
with a certain coupling $G_A$.

In principle, this term should be derived from the YM Lagrangian using
renormalization group. By an appropriate choice of $G_A$,
such a mass term may help to better quantify
lattice data. Then, if the dynamical mass is nearly constant in a certain range of
temperature,   the observed plateau in $I/T^2T_c^2$,   may  emerge even without
the dilaton potential~\cite{CMS,matrix}.
Consequently,  this approach could  be an alternative
to the formulation proposed in the previous section.

\section{Conclusions}
\label{sec:conc}

We have presented an effective theory of SU(3) Yang-Mills (YM) thermodynamics implementing the
major global symmetries, the center and scale symmetries,
and their dynamical breaking. This naturally allows a mixing between the
Polyakov loop and the dilaton field. Consequently, the magnetic confinement is
effectively embedded and results in deviations of the EoS from their
Stefan-Boltzmann limit at high temperature.

Also, matching to the 3-dimensional YM theory has been proposed, which   leads to the
gluon condensate  increasing with temperature in deconfined phase.
Contrary, in the conventional treatment of the dilaton condensate,
there is a weak thermal behavior of the composite gluon in a wide range of
temperature. This suggests, that at some temperature above $T_c$, the gluon condensate exhibits
a distinct behavior on $T$. In the
present model this temperature is roughly estimated as $\sim 2.4\,T_c$,
compatible with $\sim 2\,T_c$ extracted from the spatial
string tension~\cite{Agasian}.

We have illustrated, that the above changeover of the gluon condensate,  
becomes transparent
in the interaction measure $I={\mathcal E}-3P$ normalized by $T^2T_c^2$,
rather than by $T^4$.
Adopting the matching condition (\ref{magnetic}),
the $I/T^2T_c^2$ shows a $T^2$ raise, which is dominating at high temperature.
Before reaching the matching temperature,
the thermodynamics is well described by the model for the Polyakov
loop and a nearly-constant dilaton condensate, resulting in a monotonic 
decrease of $I/T^2T_c^2$ with $T$.
Consequently, the sum of those two contributions,  yields a plateau structure
at an intermediate temperature $T/T_c \sim 2$-$4$. This qualitative behavior
of the interaction measure is consistent with the lattice findings~\cite{latYM}.
The role of the magnetic gluon turns out to be alternative to the HTL contribution.

The nature of the physical vacuum in YM theories can be captured by topological
objects,
such as magnetic monopoles and vortices~\cite{topology}.
In the context of hot gluon plasma, it was shown within lattice 
simulations~\cite{nakamura} that
the magnetic component as a topological
defect affects crucially the thermodynamics in deconfined phase.

Different approaches were proposed to
deal with the magnetic aspect in the topological context~\cite{mag}.
In our effective theory, such magnetic feature can be attributed to the relevant global
symmetries embedded in the original color gauge group. It is desirable to examine to
what extent this effective theory is secure in describing the non-perturbative
feature at high temperature. A matching to the topological approaches could
yield more reliable constraints on the Lagrangian and its parameters.

Furthermore, introducing  quarks and
their coupling to gluons, in the proposed  theory,
could provide a scheme for an  effective description of QCD thermodynamics.

\subsection*{Acknowledgments}

We acknowledge stimulating discussions with Georg Bergner, Bengt Friman,
Larry McLerran and Owe Philipsen.
This work has been partly supported by the Hessian LOEWE initiative
through the Helmholtz International Center for FAIR (HIC for FAIR),
by the grants NS-7235.2010.2 (Russia)
and by the Polish Science Foundation (NCN).


\end{document}